\documentclass{emulateapj}
\usepackage{natbib}
\usepackage{lscape}

\newcommand{\eg}{e.g., }
\newcommand{\ie}{i.e., }
\newcommand{\Msun}{M_{\odot}}
\newcommand{\kms}{km~s$^{-1}$}
\newcommand{\ergs}{erg~s$^{-1}$}
\newcommand{\Fefs}{$^{56}$Fe}
\newcommand{\Cofs}{$^{56}$Co}
\newcommand{\Nifs}{$^{56}$Ni}

\newcommand{\Mej}{M_{\rm ej}}
\newcommand{\Mni}{M(^{56}{\rm Ni})}
\newcommand{\KE}{E_{\rm K}}
\newcommand{\vph}{v_{\rm ph}}
\def\ion#1#2{{\rm #1}~{\sc #2}}
\def\gsim{\mathrel{\rlap{\lower 4pt \hbox{\hskip 1pt $\sim$}}\raise 1pt
\hbox {$>$}}}
\def\lsim{\mathrel{\rlap{\lower 4pt \hbox{\hskip 1pt $\sim$}}\raise 1pt
\hbox {$<$}}}

\shorttitle{The evolution of the peculiar Type Ia  SN 2005hk}
\shortauthors{Sahu et al.} 

\begin{document}


\title{The evolution of the peculiar Type Ia supernova SN 2005hk over 400
days}

\author{
D.K. Sahu\altaffilmark{1},
Masaomi Tanaka\altaffilmark{2},
G.C. Anupama\altaffilmark{1},
Koji S. Kawabata\altaffilmark{3},
Keiichi Maeda\altaffilmark{4,5},
Nozomu  Tominaga\altaffilmark{2},
Ken'ichi Nomoto\altaffilmark{2,5,6},
Paolo A. Mazzali\altaffilmark{4,6,7,8}, and
T.P.Prabhu\altaffilmark{1} }
\email{dks@iiap.res.in, mtanaka@astron.s.u-tokyo.ac.jp}

\altaffiltext{1}{Indian Institute of Astrophysics, II Block Koramangala,
Bangalore 560034, India}
\altaffiltext{2}{Department of Astronomy, Graduate School of Science,
University of Tokyo, Hongo 7-3-1, Bunkyo-ku, Tokyo 113-0033, Japan}
\altaffiltext{3}{Hiroshima Astrophysical Science Center, Hiroshima
University, 1-3-1 Kagamiyama, Higashi-Hiroshima, Hiroshima 739-8526,
Japan}
\altaffiltext{4}{Max-Planck Institut f\"ur Astrophysik,
Karl-Schwarzschild-Strasse 1, Postfach 1317,  D-85741 Garching, Germany}
\altaffiltext{5}{Institute for the Physics and Mathematics of the
Universe, University of Tokyo, Kashiwa, Chiba 277-8582, Japan}
\altaffiltext{6}{Research Center for the Early Universe, Graduate School
of Science, University of Tokyo, Bunkyo-ku, Tokyo 113-0033, Japan}
\altaffiltext{7}{INAF-Osservatorio Astronomico di Trieste, via Tiepolo 11,
I-34131, Trieste, Italy}
\altaffiltext{8}{Kavli Institute for Theoretical Physics, University of California,
Santa Barbara, CA 93106}
\begin{abstract}
$UBVRI$ photometry and medium resolution optical spectroscopy of peculiar
Type Ia supernova SN 2005hk are presented and analysed, covering the
pre-maximum phase to around 400 days after explosion. The supernova is
found to be underluminous compared to "normal" Type Ia supernovae. The
photometric and spectroscopic evolution of SN 2005hk is remarkably similar to
the peculiar Type Ia event SN 2002cx. The expansion velocity of the supernova
ejecta is found to be lower than  normal Type Ia events. The  spectra
obtained $\gsim  200$ days since explosion do not show the presence of forbidden 
[\ion{Fe}{ii}], [\ion{Fe}{iii}] and [\ion{Co}{iii}] lines, but
are dominated by narrow, permitted \ion{Fe}{ii}, NIR \ion{Ca}{ii} and
\ion{Na}{i} lines with P-Cygni profiles.
Thermonuclear explosion model with Chandrasekhar
mass ejecta and a kinetic energy smaller ($\KE = 0.3 \times 10^{51} 
{\rm ergs}$) than that of canonical Type Ia supernovae is found to well
explain the observed bolometric light curve.
The mass of \Nifs\ synthesized in this explosion is $0.18 \Msun$.
The early spectra are successfully modeled with this less energetic model
with some modifications of the abundance distribution.
The late spectrum is explained as a combination of a photospheric
component and a nebular component.

\end{abstract}

\keywords{supernovae: general --- supernovae: individual (SN~2005hk)}


\section{Introduction}
\label{sec:introduction}

An impressive homogeneity in the light curves and peak luminosities make
Type Ia
supernovae (SNe Ia) good candidates in the determination of the extragalactic
distance scale. Though a majority of the observed SNe Ia belong to the
"normal"   type (Branch et al.\ 1993), a number of studies indicate
significant
photometric as well as spectroscopic differences. For example, studies 
 of nearby supernovae by Li et
al.\ (2001) indicate that 64\% SNe Ia are "normal", while 20\% are of the
overluminous SN 1991T type and 16\% of the underluminous SN 1991bg type. The
peak absolute luminosities of SNe Ia are well correlated with their immediate
post-maximum decline rate, forming a photometric sequence from the
luminous blue
events with a relatively slow decline rate to the faster, red, subluminous
events (Hamuy et al.\ 1996a,b; Phillips et al.\ 1999).

However, there are a few SNe Ia that are known to deviate from this relation,
the most notable amongst them being SN 2002cx. This supernova was found to be
underluminous, but had a light curve decline rate $\Delta$m$_{15}(B)=1.29$,
comparable to normal SNe (Li et al.\ 2003). The early phase ($\lsim 100$ days
after explosion) spectra indicate line velocities lower by a factor of 2
compared to those of normal SNe Ia (Li et al.\ 2003; Branch et al.\ 2004).
Furthermore, the late phase ($\sim250$ days after explosion) spectra are also
quite dissimilar compared to normal SNe Ia and possibly consist of P-Cygni
profiles (Jha et al.\ 2006).

Interestingly, SN 2002cx is not a unique event. Jha et al.\ (2006) and
Phillips
et al.\ (2007) have shown that SN 2005hk shows photometric and spectroscopic
behaviour almost identical to that of SN 2002cx. Further, Jha et al.\
(2006)  have
shown that SNe 2003gq and 2005P also show spectra similar to SN 2002cx.
Spectropolarimetric observations of SN 2005hk at the early phases by Chornock
et al.\ (2006) indicate low polarization levels, indicating that the
peculiarities of SN 2002cx-like SNe do not result from an extreme
asphericity.
Based on the analyses of the early phase spectra of SN 2002cx, Branch et al.\
(2004) suggest that the observed lower line velocities are consistent with
the deflagration models of explosion. Jha et al.\ (2006) report a possible
detection of \ion{O}{i} lines in the late phase spectra and suggest large
scale mixing
in the central region, which is also consistent with three-dimensional 
(3D) deflagration models.  Phillips et al.\ (2007) find that, qualitatively the observed 
light curves of SN 2005hk are in reasonable agreement 
with  model calculations of a 3D deflagaration model that produces 
$\sim 0.2~\Msun$ of $^{56}$Ni.

Understanding the nature of this class of SNe Ia is thus quite important
for the study of  homogeneity and heterogeneity of SNe Ia. It may provide a
caution for the cosmological use of SNe Ia and a strong constraint to the
explosion models.

We present in this paper the photometric and spectroscopic development of
SN 2005hk over $\sim400$ days since explosion.

\section{Observations and data reduction}
\subsection{Photometry}
Supernova SN 2005hk was observed in UBVRI bands with the 2m Himalayan Chandra
Telescope (HCT) of the Indian Astronomical Observatory (IAO), Hanle, India,
using the Himalaya Faint Object Spectrograph Camera (HFOSC), equipped with a
SITe 2k x 4k CCD. The central 2k x 2k pixels were used for imaging which
corresponds to a field of view of $10^\prime \times 10^\prime$, with a scale
of 0.296 arcsec/pixel. The monitoring of the supernova started on 2005
November
5 (JD 2,453,680), $\sim$ 4  days and $\sim$ 6  days after its discovery by Burket \&
Li (2005), and  Barentine, et al. (2005), and
continued until 2006 September 20 (JD 2,453,999). Standard fields PG0231+051,
PG1047+003, PG0942-029, PG0918+027 (Landolt 1992) were observed on 2005
November 24 and 2005 December 28 under photometric conditions, and were used
to calibrate a sequence of secondary standards in the supernova field.

\begin{figure}
\epsscale{1.0}
\plotone{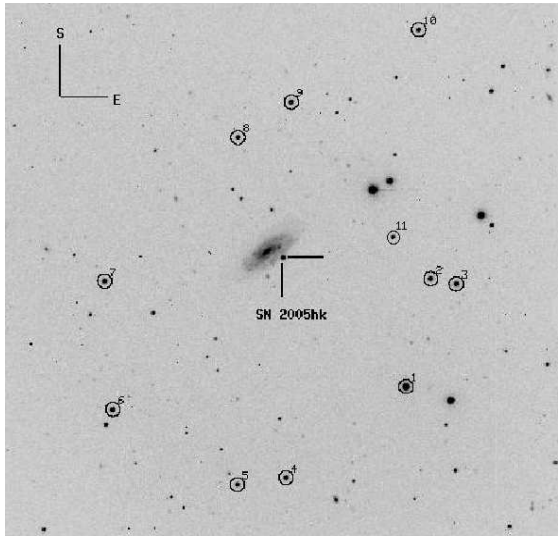}
\caption{Identification chart for SN 2005hk. The stars used as local
standards are marked as numbers 1-11.
\label{fig:field} }
\end{figure}

The data reduction was done in the standard manner, using 
various tasks available within IRAF. The observed data were bias subtracted,
flat-field corrected and cosmic-ray hits removed. Aperture photometry was
performed on the standard stars, using an aperture radius determined using
the
aperture growth curve, and were calibrated using the average color terms and
photometric zero points determined on the individual nights. The $UBVRI$
magnitudes of the secondary standards in the supernova field, calibrated and
averaged over the two nights are listed in Table \ref{tab:calmag}. The
secondary sequence is shown marked in Figure \ref{fig:field}. The magnitudes
of the supernova and the local standards were estimated using the profile
fitting technique, with a fitting radius equal to the FWHM of the stellar
profile. The difference between the aperture and profile-fitting magnitude
was
obtained using  bright standards in the supernova field and this
correction
was applied to the supernova magnitude. The calibration of the supernova
magnitude to the standard system was done differentially with respect to the
local standards.

SN 2005hk was also observed on 2005 December 26, 2006 June 30 and
November 27 with the Faint Object Camera and
Spectrograph (FOCAS; Kashikawa et al.\  2002) mounted
on the 8.2 m Subaru Telescope in the $B$ and $R$ bands.
The local standard star 11 was calibrated as indicated in
Table 1 using the HCT/HFOSC photometry of the star 2
in Figure 1, then the $B$ and $R$ magnitudes were derived by
the point spread photometry in the DAOPHOT package of IRAF.

The observed supernova magnitudes are listed in Table \ref{tab:mag}.

\subsection{Spectroscopy}
Spectroscopic monitoring of the supernova using the HCT began on 2005
November
4 and continued until 2006 January 12. The journal of observations is given
in Table \ref{tab:spec}. The HCT spectra cover the wavelength ranges
3500-7000 \AA\ and 5200-9100 \AA\ at a spectral resolution of $\sim$ 7
\AA. The
data reduction was carried out in the standard manner using the tasks
available
within IRAF. The data were bias corrected, flat-fielded and the one
dimensional
spectra extracted using the optimal extraction method. Spectra of FeAr and
FeNe lamps were used for wavelength calibration. The instrumental response
curves were obtained using spectrophotometric standards observed on the same
night and the supernova spectra were brought to a relative flux scale. The
flux
calibrated spectra in the two regions were combined to a weighted mean to
give
the final spectrum on a relative flux scale. The spectra were brought to an
absolute flux scale using zero points derived by comparing the observed flux
in the spectra with the flux estimated using the photometric magnitudes.

Spectra during the late phases were  obtained with the Subaru/FOCAS on
2005 December 26, 2006 June 30 and November 27. A 0.8 arcsec  width slit
was used
with the B300 grism, which gave a wavelength coverage of 4700-9000 \AA \
and a
spectral resolution of 11 \AA. The exposure times were 900 sec, 1,800 sec
and 3,600 sec
for 2005 Dec 26, 2006 Jun 30 and Nov 27, respectively. The instrumental
response
was corrected using data of a spectrophotometric standard star (GD 153 or
BD+28d4211). The
flux was then scaled to be consistent with the R-band photometry.

\section{$UBVRI$ Light Curves}

\begin{figure}
\epsscale{1.0}
\plotone{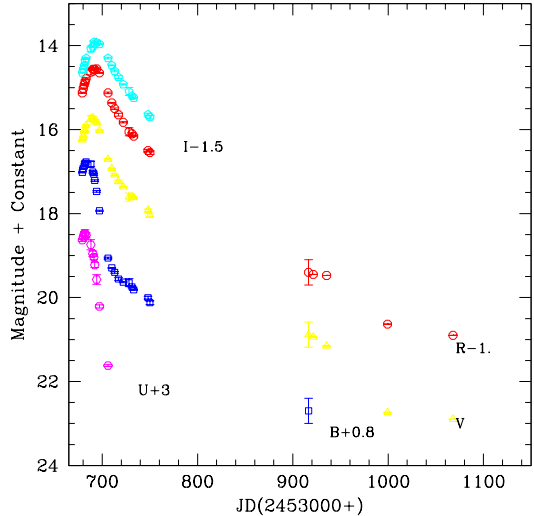}
\caption{The $UBVRI$ \  LCs of SN 2005hk. The LCs have been shifted by the
amount indicated in the legend.
\label{fig:lc} }
\end{figure}

The early phase light curves (LCs) of SN 2005hk are discussed  by 
Phillips et al.\ (2007) and Stanishev et al.\ (2006). We present
here the LCs of this supernova during the early as well as the late
phase (ref. Table \ref{tab:mag}). Figure \ref{fig:lc} shows the LCs of
SN 2005hk in the $U$, $B$, $V$, $R$ and $I$ bands, respectively. The epoch of
maximum in each of these bands is estimated by a cubic spline fit to the
points around maximum. The supernova reached a maximum in the $B$ band
on JD\,2453685.34 with magnitude of
$15.91\pm0.03$. The maximum brightness in the $V, R$ and $I$ bands occured
$\sim$  +4, +6.5 and +8.5 days, relative to the maximum in $B$ band,
respectively. The decline rate parameter in the $B$ band estimated using our
data is $\Delta$m$_{15}(B)$ = 1.68$\pm$0.05. Our estimates of the dates of
maximum, apparent maximum magnitudes and the decline rate parameter are
listed in Table \ref{tab:lcparam}. These estimates are consistent with the
values reported by Phillips et al.\ (2007) and Stanishev et al.\ (2006).

\begin{figure}
\epsscale{1.0}
\plotone{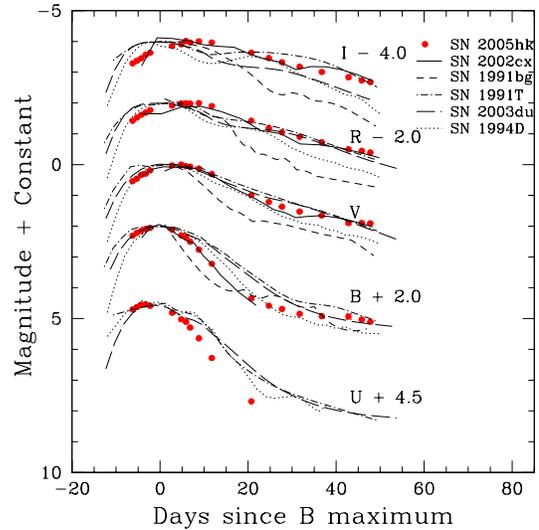}
\caption {Early phase ($< 50$~days) LC of SN 2005hk together with those of 
SNe 2002cx, 1991bg, 1991T,  2003du and  1994D. The ordinate in each panel is the
magnitude below the respective maximum, and the abscissae represent the days
since respective $B$ maximum. For clarity, the LCs have been shifted by the amount
indicated in the legend.
\label{fig:compu} }
\end{figure}

\begin{figure}
\epsscale{1.0}
\plotone{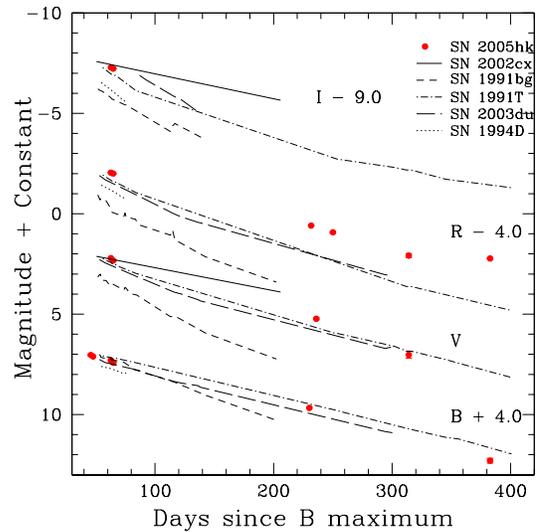}
\caption{Late phase ($>50$~days) LC of SN 2005hk together with those of 
SNe 2002cx, 1991bg, 1991T, 2003du and  1994D. The ordinate in each panel is the
magnitude below the respective maximum, and the abscissae represent the days
since respective $B$ maximum. For clarity, the LCs have been shifted by the amount
indicated in the legend.
\label{fig:compb}}
\end{figure}

The LCs of SN 2005hk are compared with those of the normal SNe 1994D
[$\Delta m_{15}(B)=1.31\pm0.08$; Richmond et al.\ 1995] and 2003du
[$\Delta m_{15}{B}=1.04\pm0.04$; Anupama et al.\ 2005], the overluminous
SN 1991T
[$\Delta m_{15}(B)=0.95\pm0.05$; Lira et al.\ 1998], the subluminous SN
1991bg
[$\Delta m_{15}(B)=1.93\pm0.08$; Filippenko et al.\ 1992; Turatto et al.\
1996]
and the peculiar SN 2002cx [$\Delta m_{15}(B)=1.29\pm0.011$;
Li et al.\ 2003, $\Delta m_{15}(B)=1.7\pm0.1$; Phillips et al.\ 2007] (Figures
\ref{fig:compu}--\ref{fig:compb}).
The early LCs (Figure \ref{fig:compu}) indicate that the pre-maximum brightening of
SN 2005hk was faster than the luminous SN 1991T but comparable to the normal
SNe 1994D and 2003du. Following the maximum, during the initial 20 days, the
decline in $B$ was faster than in normal SNe, but similar to SN 2002cx,
 while the decline in $V$ was similar to the normal SN 1994D.
Subsequently, beyond 30 days past maximum, the decline in $B$ and $V$ was
slower than all the other SNe. The secondary peak seen in the $R$ and $I$
bands of SNe Ia was found to be absent in SN 2005hk, similar to the
subluminous SN 1991bg. However, beyond $\sim$ 50 days past maximum the $R$ LC
of SN 2005hk had a decline similar to SN 1991T, which is slower than that
seen in the normal and subluminous SNe (Figure \ref{fig:compb}). The evolution of 
the $U$ light curve
of SN 2005hk was found to be faster than both the normal and the luminous
types.

 The early phase evolution of SN 2005hk is very similar to that of SN 2002cx.
Both objects have  similar $\Delta$m$_{15}(B)$. SN 2005hk has a 
$\Delta$m$_{15}(B)$ of $1.68$, while SN 2002cx has a value of 1.7, as
indicated by the the revised photometry by Phillips et al.\ (2007). 
On the other hand, at
epochs beyond 15 days past maximum, SN 2005hk declines somewhat slower
than SN 2002cx. The decline rate during $\sim 20-50$ days past $B$
maximum, in the $B$, $V$, $R$ and $I$ bands are, respectively, 0.021, 0.027,
0.031 and 0.031 mag day$^{-1}$, while the corresponding decline rates in
SN 2002cx during the same phase are 0.037, 0.035, 0.041 and 0.035. It may be
noted here that no galaxy template subtraction is done for the data
presented here, and the SN magnitudes may be affected by the host galaxy at
the later phases. However, as suggested by Phillips et
al.(2007), neither $K$ correction nor the galaxy background contamination is
sufficient to explain the discrepancies seen in the LCs of SN 2005hk and
SN 2002cx.

\section{Reddening and Color curves}

The reddening within our Galaxy in the direction of SN 2005hk is
$E(B-V )_{\rm{Gal}}=0.022$ (Schlegel et al.\ 1998). The reddening within the
host galaxy may be estimated from the methods suggested by Phillips
et al.\ (1999), Altavilla et al.\ (2004) and Lira\ (1996). The Phillips
et al.\ (1999) and the Altavilla et al.\ (2004) methods give
$E(B-V)_{\rm{Host}}$ = 0.20
and 0.09, respectively, while the Lira\  (1996) method indicates a higher
value.
Spectropolarimetric observations of SN 2005hk (Chornock et al.\  2006)
indicate an interstellar polarization of 0.27\%
produced by the host galaxy which, for
the standard dust polarization efficiency, corresponds to
$E(B-V)_{\rm{Host}} = 0.09$. Owing to the peculiar nature of SN 2005hk, the
methods based on normal SNe Ia may not be applicable. Hence, adopting the
estimate for the host galaxy extinction as indicated by the
spectropolarimetric observations, we estimate the total
reddening towards SN 2005hk to be $E(B-V)_{\rm{tot}} = 0.11$.

\begin{figure}
\epsscale{1.0}
\plotone{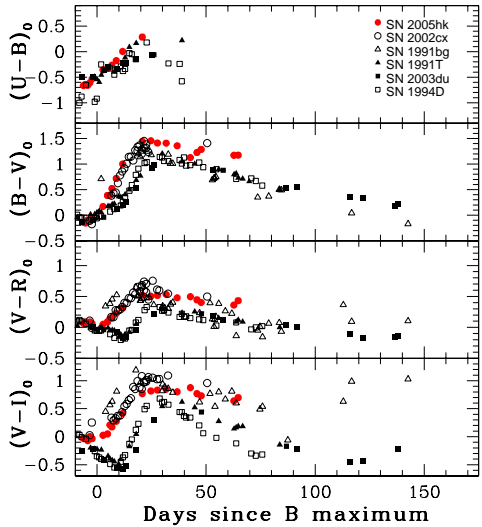}
\caption{The $U-B$, $B-V$, $V-R$ and $V-I$ color curves of SN 2005hk. Also
shown for comparison are the color curves of SNe 2002cx, 1991T, 1991bg, 2003du and
1994D. The abscissae correspond to days since respective $B$ maximum.
\label{fig:color}}
\end{figure}

The $(U-B)$, $(B-V)$, $(V-R)$ and $(V-I)$ color curves of SN 2005hk are
shown
in Figure \ref{fig:color}, compared with other SNe. The color curves of
all SNe have been
corrected for reddening using the Cardelli et al.\ (1989) extinction law and
the $E(B-V)$ values of  $E(B-V)$ = 0.13 for
SN 1991T (Phillips et al.\ 1992), $E(B-V)$ = 0.034 for SN 2002cx (Li et
al.\ 2003),
$E(B-V)$ = 0.04  for SN 1994D (Richmond et al.\ 1995), $E(B-V)$ = 0.05 for
SN 1991bg  (Turatto et al.\ 1996) and $E(B-V)$ = 0.01 for SN 2003du (Anupama
et al.\ 2005).

The color curves of SN 2005hk are very similar to SN 2002cx. The $(U-B)$
color
evolution is very similar to SN 1991T. The $(B-V)$ and $(V-R)$ color
evolution
of SN 2005hk in the pre-maximum and maximum phase are very similar to the
other
SNe Ia except SN 1991bg, which had a considerably red color. Beyond $\sim 5$
days, until $\sim 20$ days after $B$ maximum, both $(B-V)$ and $(V-R)$
colors get increasingly redder compared to other SNe, but are still bluer
compared to SN 1991bg. The $(V-I)$ color on the other hand gets redder than 
other SNe $\sim$ 5 days before maximum, and also does not show the dip seen in 
the color curves of other SNe Ia at $\sim$ 10 days post-maximum. Beyond 
$\sim$20 days after maximum, the $(V-R)$ and $(V-I)$ colors of SN 2005hk 
follow the trend of other supernovae and become marginally blue, while 
$(B-V)$ continues to be redder than other SNe, including SN 1991bg.

Interestingly, despite a very similar color evolution during the pre-maximum
phase to $\sim 20$ days after $B$ maximum, SN 2005hk appears to be marginally
bluer than SN 2002cx at later phases.

\section{Bolometric light curve}

\subsection{Behavior of bolometric light curve}

The bolometric LC of SN 2005hk is estimated using the $UBVRI$
photometry  presented here along with the NIR $YJH$ photometry reported by
Phillips et al.\ (2007). The magnitudes were reddening corrected using the
value estimated in the previous section and the Cardelli et al.\ (1989) 
extinction law. The corrected magnitudes were then converted to monochromatic 
fluxes using the zero points from Bessell et al.\ (1998). The bolometric 
fluxes were derived by fitting a spline curve to the $U, B, V, R, I, Y, J$ and 
$H$ fluxes and integrating over the wavelength range 3100\AA \ to 
$1.63\micron$. In the later phases when only $B, V, R$ or $V, R$
magnitudes were available, the bolometric flux were estimated by applying
a bolometric correction to the available magnitudes. Since SN 2005hk belongs
to a rare class of SNe Ia, the bolometric correction is derived from the
light curve of SN 2005hk itself rather than assuming the corrections based on
normal SNe Ia (Suntzeff\ 1996; Contrado et al.\ 2000). 
The bolometric 
corrections were estimated based on the last three points in the LC for which 
the $BVRIJH$ bolometric flux is available. A simple average of the bolometric
corrections for these days give values of $B.C = -0.926\pm0.089$ in $B$,
$B.C=0.294\pm0.045$ in $V$ and $B.C=0.677\pm0.027$ in $R$ band. 

\begin{figure}
\epsscale{1.0}
\plotone{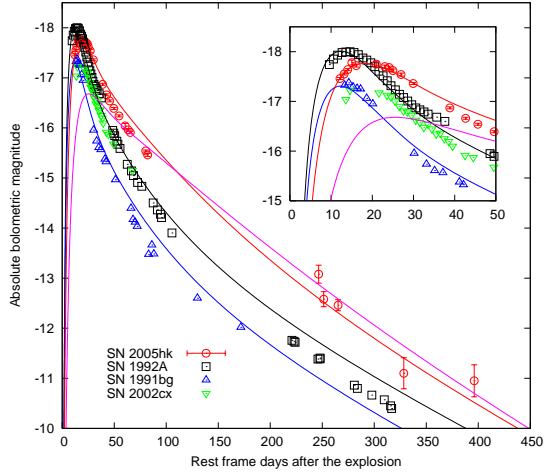}
\caption{
 Bolometric LC of SN 2005hk (red circles) compared with a synthetic LC
with the E03 model (red line).
Black squares and blue triangles are bolometric LCs of
SNe 1992A and 1991bg (Suntzeff 1996), respectively.
Model LCs for these two SNe are shown in black and blue lines, respectively
(based on W7 but different mass of \Nifs, see Table \ref{tab:param}).
Green triangles are the bolometric LC of SN 2002cx.
In order to fix the rise time of the LCs of SNe 1992A and 1991bg,
$B$ band rise time is assumed as $t_{rise} = 19.5 \times s$ days
(Conley et al. 2006),
where $s$ is stretch factor of $B$ band LC (Goldhaber et al. 2001).
The $B$ band rise times of SNe 1992A ($s = 0.80$) and 1991bg ($s = 0.68$)
are estimated as 15.6 and 13.2 days, respectively.
Since the LC of SN 2005hk is deviated from the normal sequence of SNe Ia,
we simply assume  $t_{rise} = 17$ days as in Phillips et al. (2007).
 The magenta line shows the synthetic LC with the E008 model.
\label{fig:modelLC}}
\end{figure}

The bolometric LC is plotted in Figure \ref{fig:modelLC} (red circles). 
Also plotted in the figure are the bolometric LCs of 
the normal (but slightly underluminous) Ia SN 1992A (black squares), 
the subluminous Ia SN 1991bg (blue triangles) and SN 2002cx (green triangles).
The peak luminosity ($M_{bol} = -17.7$) is fainter than the canonical 
value for normal SNe Ia.
It suggests that a smaller amount of \Nifs\ is
synthesized during the explosion. 

 The decline of the bolometric LC
after the maximum is slower than that of SN 1992A
although they have a similar maximum luminosity, 
as also shown by Phillips et al. (2007).
Our observations extend the bolometric LC to $\sim 400$ days since explosion. 
At such late epochs, the bolometric luminosity of SN 2005hk is
still brighter than that of SN 1992A.
In addition, the difference in the luminosity between the two SNe
becomes large, reaching $\sim 1$ magnitude at $> 250$ days after
the explosion. The bright late phase luminosity indicates a more efficient
trapping of the $\gamma$-rays from decaying \Cofs\ in SN 2005hk compared to
SN 1992A. 

\subsection{Light curve modeling}

Phillips et al. (2007) suggest that qualitatively the observed properties of SN 2005hk 
are consistent  with 3D deflagration model 
(\eg R\"opke et al. 2006, Blinnikov et al.\ 2006), showing the bolometric 
and multi-color LC model until $\sim 80$ days after the explosion.
Branch et al. (2004) and Jha et al. (2006) also suggest
a 3D deflagration model for SN 2002cx. Here we give the first investigation of 
the bolometric LC of this class of SNe until $\sim 400$ days after the 
explosion.

We use an LTE radiation transfer code (Iwamoto et al.\ 2000)
for the computation of the bolometric LCs.
For the $\gamma$-ray transport, a gray atmosphere is assumed. 
For optical radiation transport, electron-scattering and line opacity 
are taken into account.  The electron-scattering opacity is evaluated 
by solving the Saha equation 
while the line opacity is crudely assumed to be constant 
($0.1\ {\rm cm^2 \ g^{-1}}$; Iwamoto 1997) for simplicity.
We use the W7 deflagration model (Nomoto et al.\ 1984) as a 
standard density structure and abundance distributions.

First, the mass of \Nifs\ [$M$(\Nifs)] is varied from the W7 value
($M$(\Nifs) $= 0.59 \Msun$) to obtain a good agreement with the maximum 
luminosity of SNe Ia shown in Figure \ref{fig:modelLC}.
In this procedure, the total mass of iron group elements is kept 
constant to conserve the explosion energy. The black and blue lines in Figure 
\ref{fig:modelLC} are the synthetic LCs with $\Mni = 0.17$ and $0.088 \Msun$, 
respectively. 
Both at the early phase (ref.\ inset in Fig. \ref{fig:modelLC}) 
and the late phase, the LCs are in good agreement with SNe 1992A and  1991bg 
(Suntzeff\ 1996), respectively
\footnote{ This shows  that our crude assumption of constant line opacity
($0.1\ {\rm cm^2 \ g^{-1}}$) is reasonable.}. 
However, the model sequence with various $\Mni$ never 
reproduces the brightness of SN 2005hk at $\gsim 50$ days after 
the explosion.

 As also suggested by Phillips et al. (2007),
we now try to explain the LC of SN 2005hk with less energetic models
to fit the LC around maximum.
We construct a less energetic model simply 
by scaling the structure of W7 homologously. 
The mass of the burned material should be 
adjusted accordingly by taking nuclear energy production into account.
In our model, the mass fraction of each burned element is reduced 
by the same fraction
[$(\KE + E_{\rm B} )/(E_{\rm W7} + E_{\rm B})$] in every radial shell. 
Here $\KE$, $E_{\rm B}$ and $E_{\rm W7}$ are 
the kinetic energy of the SN ejecta, the binding energy of a progenitor WD 
($\sim 0.5 \times 10^{51}$ ergs), and the kinetic energy of W7 model 
($\sim 1.3 \times 10^{51}$ ergs), respectively. 
The mass of the burned elements reduced by the procedure above
is compensated by unburned C and O, with the mass fraction of
X(C)=0.5 and X(O)=0.5.

\begin{figure}
\epsscale{1.0}
\plotone{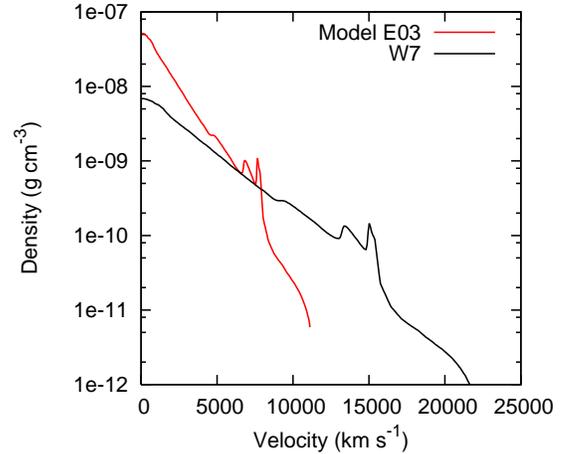}
\caption{
The density structure of model E03 compared with that of W7
(1 day after the explosion).
The kinetic energy of model E03 is $\KE = 0.3 \times 10^{51} {\rm ergs}$.
The density structure is scaled from that of W7 homologously.
This makes the density at the inner part higher than the original.
\label{fig:dens}}
\end{figure}

The red line in Figure \ref{fig:modelLC} is the synthetic LC with 
$\Mni = 0.18 \Msun$ and $\KE = 0.3 \times 10^{51} {\rm ergs}$. 
This kinetic energy is similar to that 
of the 3D deflagration model used in Phillips et al.\ (2007).
 The density structure of this model is given in Figure \ref{fig:dens}
together with that of the original W7.
This model gives a reasonable fit 
to SN 2005hk at late phase as well as at early phase.
 In the figure, we assume $B$ maximum of SN 2005hk occurs 
17 days after the explosion, which gives the best fit of the LC
(see also Phillips et al. 2007 
\footnote{The rise time of $15 \pm 1$ days is estimated by 
Phillips et al. (2007), using $3\sigma$ upper limit 
at 15 days before B maximum.  
Our model LC assuming the rise time of 17 days is consistent 
with the upper limit.}).
The mass of \Nifs\ derived from the model is somewhat smaller  than
$0.24 \Msun$ derived by Phillips et al. (2007).
It may be caused by a different distribution of \Nifs\ in the ejecta.
Since \Nifs\ in the model in Phillips et al. (2007) is extended
to the outer layers, a part of \Nifs\ does not contribute to the optical light.
In this paper, we call our scaled, less energetic model E03.

 The behavior of the model E03 LC is understood by simple analysis.
If the opacity is assumed to be constant, 
the time scale of the bolometric LC  ($\tau_{LC}$) is scaled as
$\tau_{\rm LC} \propto \Mej^{3/4} \KE^{-1/4}$ (Arnett 1982), 
where $\Mej$ is the mass of the ejecta.
Therefore, if $\Mej$ is fixed (Chandrasekhar mass), 
the lower energy gives a slow decline after the maximum.
The late phase evolution of LC is mainly determined by the 
optical depth to $\gamma$-rays.
Since the optical depth is higher for lower energetic ejecta
($\tau \propto \Mej^2 \KE^{-1}$, \eg Maeda et al. 2003),
the LC of model E03 declines slowly at late phase.

The characteristic parameters of the LC models are summarized 
in Table \ref{tab:param}. 
In the lower energetic model, unburned C and O should be 
abundant, being $\sim 0.8 \Msun$ in total. 
 The presence of large amount of unburned C+O
is also seen in 3D deflagration model (R\"opke et al. 2006), 
which is used in Phillips et al. (2007).
Although our model is very simple, it mimics the properties
of the  3D deflagration model.
Our model is further verified by spectral modeling in \S 6.3.

\section{The spectrum}
\subsection{Evolution during early phases}

\begin{figure}
\epsscale{1.0}
\plotone{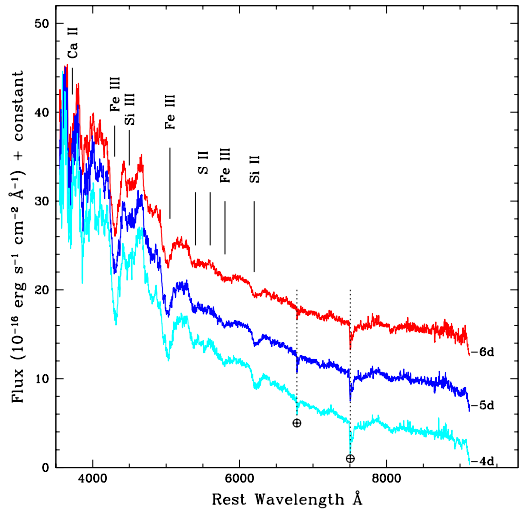}
\caption{Pre-maximum spectral evolution  of SN 2005hk. The phases marked
are relative to date of $B$ maximum. The spectra have  been corrected for
the redshift of the host galaxy, but not corrected for the reddening. For
clarity the spectra have been shifted vertically. The telluric lines have 
not been
removed from the spectra and are marked with the symbol {\small $\oplus$}.
\label{fig:spec1}}
\end{figure}

\begin{figure}
\epsscale{1.0}
\plotone{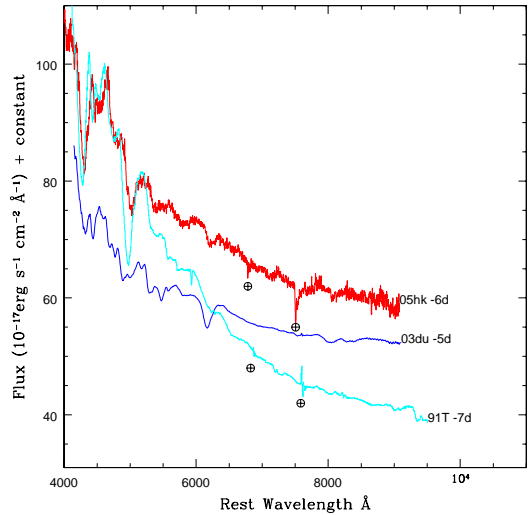}
\caption{Pre-maximum spectrum of SN 2005hk compared with those of 
Type Ia SNe 1991T and  2003du at similar epoch.  The phases marked are
relative to date of $B$ maximum. The spectra have  been corrected for
the redshift of the host galaxy, but not corrected for the reddening.
For clarity the spectra have been shifted vertically. The telluric lines 
have not been removed from the spectra of SNe 2005hk and  1991T, they are 
marked with  the symbol  {\small $\oplus$}.
\label{fig:comp_photo_03duu1}}
\end{figure}

\begin{figure}
\epsscale{1.0}
\plotone{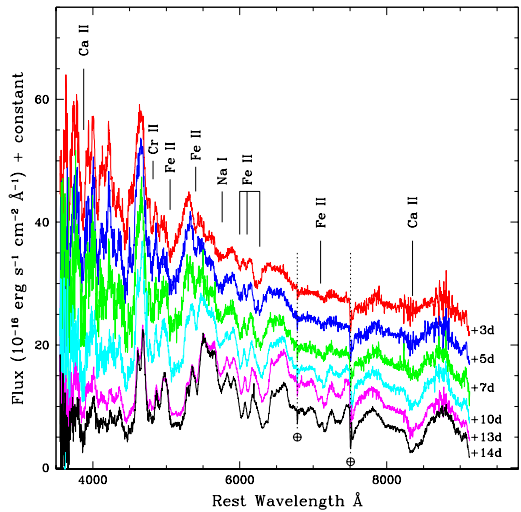}
\caption{Spectral evolution  of SN 2005hk from phase $+3$ days to
$+14$ days relative to date of $B$ maximum. The spectra have  been
corrected for the redshift of the host galaxy, but not corrected for
the reddening. For clarity the spectra have been shifted vertically.
The telluric lines 
have not been removed from the spectra  and are marked with 
the symbol {\small $\oplus$}. Line identification is based on
Branch et al. (2004)
\label{fig:spec2}}
\end{figure}

\begin{figure}
\epsscale{1.0}
\plotone{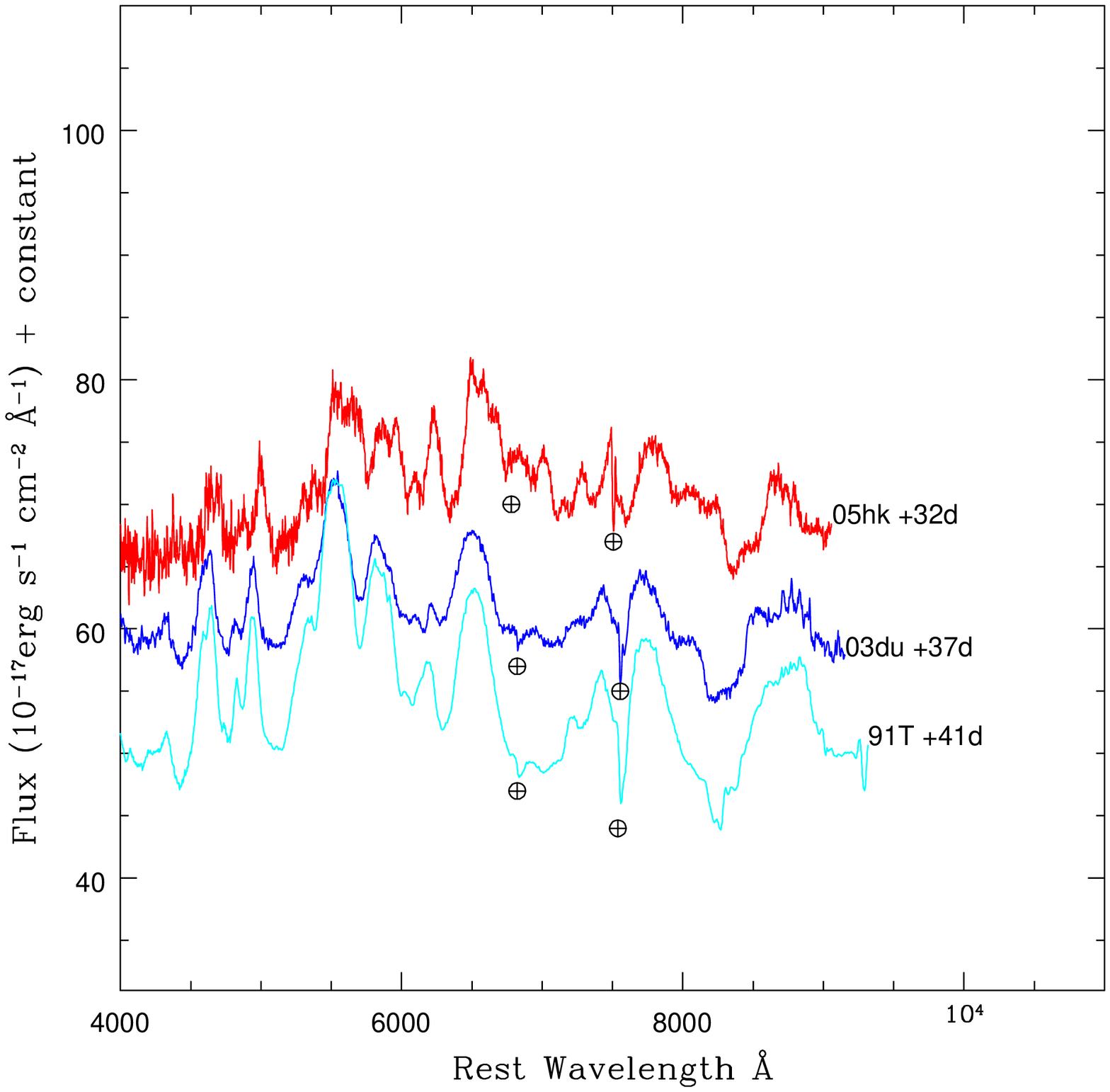}
\caption{Post-maximum spectrum   of SN 2005hk compared with those
of Type Ia SNe 1991T and  2003du at similar epochs.  The phases
marked are relative to date of $B$ maximum. The spectra have  been
corrected for the redshift of the host galaxy, but not corrected for
the reddening. For clarity the spectra have been shifted vertically.
The telluric lines  have not been removed from these spectra 
and are marked with the symbol {\small $\oplus$}.
\label{fig:comp_photo_03duu2}}
\end{figure}

\begin{figure}
\epsscale{1.0}
\plotone{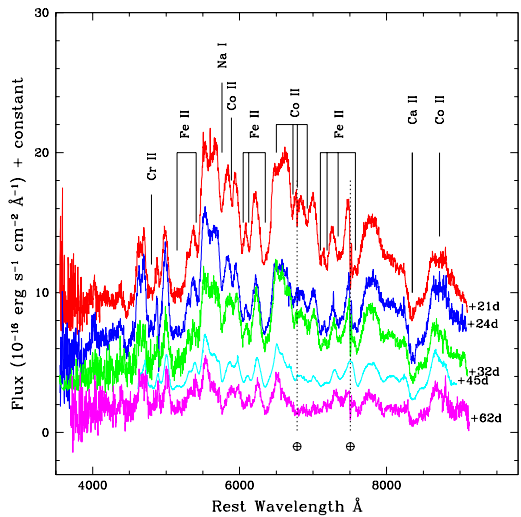}
\caption{Spectral evolution  of SN 2005hk from phase $+21$ days to $+62$ days
relative to date of $B$ maximum. The spectra have been corrected for the
redshift of the host galaxy, but not corrected for the reddening. For clarity
the spectra have been shifted vertically.  The telluric lines 
have not been removed from the spectra of SN 2005hk taken with HCT and are 
marked with the symbol {\small $\oplus$}. The line identification is based on
Branch et al. (2004).
\label{fig:spec3}}
\end{figure}

The spectral evolution of SN 2005hk during the early phases are shown in
Figure \ref{fig:spec1} -- Figure \ref{fig:spec3}. The pre-maximum spectra 
(ref. Fig.\ref{fig:spec1}) show a blue continuum, dominated by \ion{Fe}{iii}
lines, with weak \ion{Si}{ii} and \ion{Ca}{ii} H\&K absorption (see \S 6.3.1).
Comparing the pre-maximum spectrum with other normal Type Ia as well as 
peculiar Type Ia (Fig. \ref{fig:comp_photo_03duu1}), it is seen that the 
spectrum of SN 2005hk is very similar to that of SN 1991T. However, the minima 
of the \ion{Ca}{ii} H\& K, \ion{Fe}{ii}, \ion{Fe}{iii} and \ion{Si}{ii} lines,
indicate that SN 2005hk has much lower expansion velocities that decrease from 
$\sim 6900$ km s$^{-1}$ on day $-6$ to $\sim 6200$ km s$^{-1}$ on day $-4$. The
similarity of the pre-maximum spectrum of SN 2005hk with SN 1991T-like events
has also been noted by Phillips et al.\ (2007) \& Stanishev et al.\ (2006).

The post-maximum evolution of the spectrum of SN 2005hk is presented in
Figures \ref{fig:spec2} and \ref{fig:spec3}. The spectrum and its evolution 
are quite different from normal SNe Ia, as can be seen from Figure
\ref{fig:comp_photo_03duu2}. The spectrum and its evolution closely matches
that of SN 2002cx (Li et al.\ 2003, Branch et al.\ 2004, Jha et al.\ 2006).
The spectrum is dominated by \ion{Fe}{ii} and lines due to \ion{Co}{ii},
\ion{Na}{i} and \ion{Ca}{ii} are also clearly present (see \S 6.3.1).
Some \ion{Cr}{ii} lines may contribute to the absorption around 4800 \AA\
(Branch et al.\ 2004). The photospheric velocity decreases to 
$\sim 4000$ km s$^{-1}$ by day $+10$. The flux in the blue continuum is found 
to decline steadily, and by day $+14$, the continuum is quite weak below 
4500\AA. This could also be an effect of line blanketing due to \ion{Fe}{ii} 
lines. The spectrum remains almost unchanged from day $+10$ to day $+24$.

The spectrum of day $+62$ is very similar to that of day $+45$, except for the
narrowing of lines (ref. Fig.\ref{fig:spec3}). 
The lines in the 6500-6800 \AA, identified with \ion{Co}{ii} 
(Branch et al. 2004), which by day $+32$ started weakening steadily with time
(but they could be \ion{C}{ii}, see \S 6.3.1). 

\subsection{The late phases}

\begin{figure}
\epsscale{1.0}
\plotone{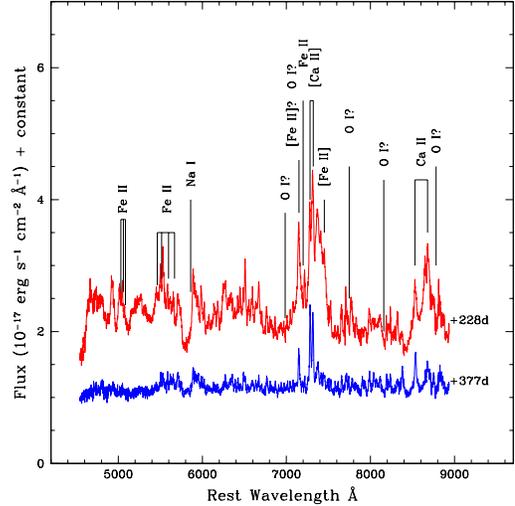}
\caption{Late phase spectrum of SN 2005hk at $+228$ days and $+377$ days
relative to date of $B$ maximum. The spectra have  been corrected for
the redshift of the host galaxy, but not corrected for the reddening.
For clarity the spectra have been shifted vertically. The telluric 
lines have been removed from the spectrum.
The line identification is based on Jha et al. (2006).
\label{fig:spec4}}
\end{figure}

\begin{figure}
\epsscale{1.0}
\plotone{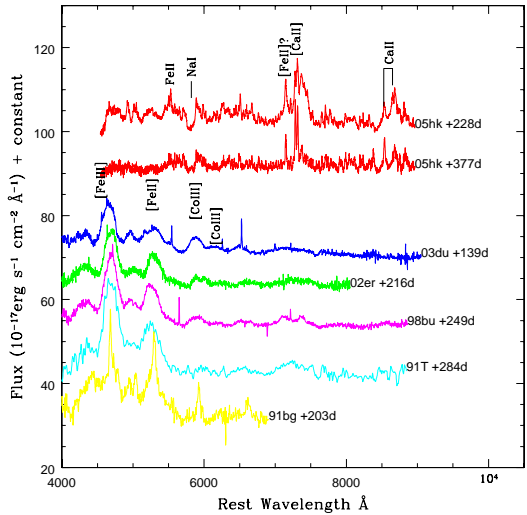}
\caption{Late phase spectrum of SN 2005hk at $+228$ days and $+377$ days
relative to date of $B$ maximum compared with those of other Type Ia SNe 2003du, 2002er, 1998bu, 1991T and 1991bg.
The spectra have  been corrected for the redshift of the host galaxy, but
not corrected for the reddening. For clarity the spectra have been shifted
vertically.
\label{fig:comp_nebular}}
\end{figure}

The spectra of SN 2005hk were obtained during the late phase on days 
$+228$ and $+377$ (Fig. \ref{fig:spec4}). As in the case of the early phase spectra, the  
spectrum of SN 2005hk during these phases is very different from the spectra 
of normal SNe Ia at similar phases.
The spectrum of SN 2005hk is compared with those of other normal and peculiar
Type Ia supernovae in Figure \ref{fig:comp_nebular}. The difference between the
spectrum of SN 2005hk and spectra of other Type Ia supernovae is obvious.
While the other SNe Ia show blends of strong forbidden lines due to iron
[\ion{Fe}{ii}] (5159\AA), [\ion{Fe}{iii}] (4500-4800 \AA) and forbidden lines 
due to cobalt [\ion{Co}{iii}] (5890\AA) (Kuchner et al.\ 1994) 
 in their late phase spectrum, no 
signature of these lines is seen in the spectrum of SN 2005hk.

The late time spectrum of SN 2005hk is dominated by \ion{Fe}{ii} lines. The 
\ion{Ca}{ii} NIR lines are  strong and \ion{Na}{i} is also present. The 
spectra of days $+228$ and $+377$ (ref. Fig.\ref{fig:spec4}) also show 
forbidden lines due to [\ion{Ca}{ii}] 7291, 7234 \AA\ and  
[\ion{Fe}{ii}] at 7155 \AA\ and 7453 \AA. The ratio of the fluxes of 
\ion{Ca}{ii} NIR and the [\ion{Ca}{ii}] lines on day $+228$ indicate densities 
of $\sim 10^9$ cm$^{-3}$ in the line emitting region. By day $+377$, the 
[\ion{Ca}{ii}] lines are stronger than the \ion{Ca}{ii} NIR lines, and the 
line ratios imply temperatures $\la 4500$~K and density  
$\sim 10^8$ cm$^{-3}$ (Fransson \& Chevalier\ 1989). The mass ejected during 
the explosion can also be estimated using the number density arrived at with 
the Ca line ratios. Assuming that the ejecta is moving with a constant velocity
of 1000 km s$^{-1}$ over a period of $\sim$ 380 days after the explosion and
a mean atomic weight $A = 40$, the estimated mass of the ejecta is
$\sim 0.25 M_\odot$. As pointed out by Jha et al.\ (2006) the mass of the
ejecta arrived at is indicative as it does not take into account factors such
as clumping or more complicated structures in the line emitting regions.
It may however be noted that a high concentration of the mass in the central
part is also suggested by the spectral modeling (\S 6.3).

The shift in the central wavelength of the [\ion{Ca}{ii}]  forbidden emission
lines indicates an
expansion velocity of $\sim 500$ km s$^{-1}$ on day $+228$ and
$\sim 300$ km s$^{-1}$ on day $+377$. The full width half maximum (FWHM) of
the emission lines indicate a velocity dispersion of $\sim 1000$ km s$^{-1}$
on day $+228$ and $\sim 500$ km s$^{-1}$ on day $+377$. Similarly, the
 emission component of \ion{Ca}{ii} NIR 8542 \AA\ line indicates an expansion
velocity of $\sim
235$ km s$^{-1}$ on day $+228$ and $\sim 210$ km s$^{-1}$ on day $+377$, with 
velocity dispersions $\sim 2100$ km s$^{-1}$ and $\sim 1200$ km s$^{-1}$,
respectively. Low velocity \ion{O}{i} at 7002 \AA\ and 7773 \AA \ absorption 
features are clearly identified in the late phase spectrum. These features were 
identified, for the first time in any SNe Ia, by Jha et al.\ (2006) in the 
''nebular'' spectrum of SN 2002cx. The velocity of these lines, measured at the 
minimum is $\sim$ 800 km s$^{-1}$. Though there is an overall similarity 
between the spectra of SN 2002cx and SN 2005hk at late phases, there are some 
noteworthy differences also. Comparing the spectrum of SN 2005hk on day $+228$ 
with the $+227$ day spectrum of SN 2002cx (Jha et al.\ 2006), it is found that 
the expansion velocities and the velocity dispersions are higher in SN 2005hk. 
The velocity dispersion based on the [\ion{Ca}{ii}] and the \ion{Ca}{ii} lines 
are in fact almost twice that estimated in SN 2002cx.

\subsection{Spectral modeling}
\subsubsection{Early Phases}
\label{sec:spearly}

The observed spectra of SN 2005hk are further investigated 
by spectral modeling.
For this purpose, we compute synthetic spectra using the Monte Carlo spectrum 
synthesis code developed by Mazzali \& Lucy\ (1993), Lucy\ (1999) and Mazzali\ 
(2000). We create model spectra at 5 epochs and compare them with the observed 
spectra ($-6$, $+3$, $+14$, $+24$ and $+45$ days from $B$ maximum).
 The rise time is assumed to be 17 days, as in the LC modeling (\S 5.2).

The Monte Carlo spectrum synthesis code assumes a spherical, sharply defined 
photosphere in the ejecta as an inner boundary. By tracing rays of a number of 
photon packets emitted from the photosphere, the temperature structure in the 
SN atmosphere is computed through the flux at each radial point (see eq. (1) 
and (2) in Mazzali \& Lucy 1993).
Radiative equilibrium is assumed for the 
temperature determination. For the interaction between photons and the SN 
ejecta, electron scattering and line scattering are taken into account. 
Sobolev approximation is used for line scattering. 
Sobolev optical depth is evaluated 
by computing ionization fractions and electron populations of each ion for a 
given temperature structure. 
Details of the physical assumptions adopted in the 
code are summarized in Mazzali \& Lucy\ (1993), Lucy\ (1999) and Mazzali\ 
(2000).

The input parameters of the code are the position 
(\ie velocity, thanks to the homologous expansion) 
of the photosphere ($v_{\rm ph}$) and the bolometric luminosity ($L$). 
The parameter $L$ is the emergent luminosity.
Since some photons emitted from the photosphere 
are back scattered into the photosphere, 
the luminosity at the photosphere 
(total photon luminosity emitted from the photosphere), 
is higher than the input bolometric luminosity $L$
(Mazzali \& Lucy 1993).
As a density structure we use the less energetic model that 
gives the best agreement with the observed LC 
(model E03, see \S 5 and Table 5).

\begin{figure}
\epsscale{1.0}
\plotone{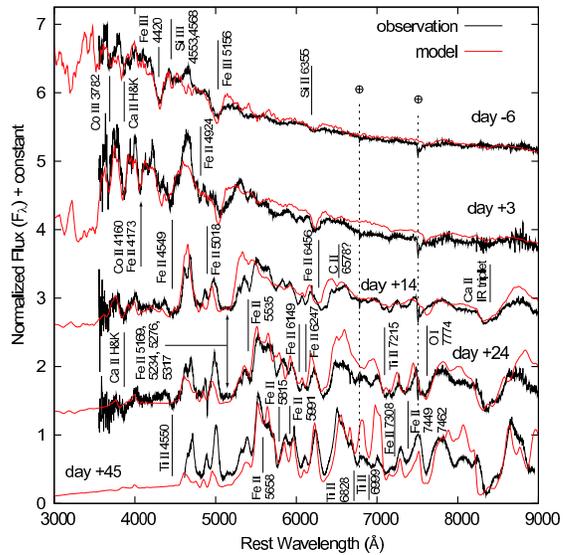}
\figcaption{
Comparison between the observed early phase spectra of SN 2005hk at
($-6$, $+3$, $+14$, $+24$ and $+45$ days from $B$ maximum,
from top to bottom) and the synthetic spectra.
Lines that make a major contribution to the absorptions are marked.
\label{fig:early}}
\end{figure}

 The mass fractions of elements are also required for the computation
of the synthetic spectra.
We treat the mass fractions as parameters
rather than using the abundance distribution in the original LC model.
The mass fractions of elements are assumed to be homogeneous
above the photosphere.
We determine the mass fraction of each element 
to give a good fit of the observed line depths.
As time goes by, the photosphere recedes
and the absorption lines are formed in the inner part of the ejecta. 
Nevertheless, we always assume a homogeneous abundance distribution
at each epoch, although we change the mass fractions of the elements.
The mass fractions derived by the modeling of each spectrum are 
thought to be a representative abundance ratio near the photosphere
because the absorption lines are formed near the photosphere
most efficiently.

The parameters in the best fitting cases are summarized 
in Table \ref{tab:spparam}.
Figure \ref{fig:early} shows the comparison between observed spectra and 
synthetic spectra. Synthetic spectra are reddened with 
$E(B-V)_{\rm Gal} = 0.022$ and $E(B-V)_{\rm Host} = 0.09$ (\S 4.). The 
distance modulus $\mu = 33.46$ is assumed. In the figure, the flux and models
 at epoch are scaled and shifted for clarity.

The spectrum of SN 2005hk at day $-6$ is in good agreement with a synthetic 
spectrum computed with $\log L$ (\ergs) $=42.56$ 
and $v_{\rm ph}$ = $6500$ \kms.
The strong absorption features around 4300\AA\ and 5000\AA\ are mainly due to 
\ion{Fe}{iii} (Mazzali et al. 1995). The photospheric temperature is computed 
as $T = 17000$ K by taking into account the back scattering effect (Mazzali \& 
Lucy\ 1993). This high temperature makes the \ion{Fe}{iii} lines strong and 
also reproduces the color of the observed spectrum.

The temperature at the photosphere is relatively higher than that of 
normal Type Ia SNe at similar epoch, and it is comparable to
the temperature at very early phase ($\sim -10$ days; Tanaka et al. 2007).
 The weakness of \ion{Si}{ii} suggests 
that the outermost ejecta of SN 2005hk do not contain much Si 
(X(Si) $\sim$ 0.02) as in normal SNe Ia (X(Si) $\gsim$ 0.5). The \ion{S}{i} 
lines are similarly weak. The mass fraction of S 
is estimated as X(S) $\sim 0.02$.
The \ion{Ca}{ii} lines are also very weak and there is no clear evidence of 
high velocity absorptions that is usually seen at pre-maximum epochs
(Mazzali et al. 2005; Tanaka et al. 2006). No C lines are seen, which gives an 
upper limit of the C mass fraction X(C) $\lsim 0.001$. The rest of the ejecta 
consist of oxygen, whose mass fraction is estimated as X(O) $\sim 0.86$.

The day $+ 3$ spectrum matches a synthetic spectrum with 
$\log (L)$ \ergs $= 42.66$ and $v_{\rm ph} =6000$ \kms. The temperature at the 
photosphere decreases ($T \sim 9,000$K) from pre-maximum epochs. The low 
temperature allows some singly ionized ions such as \ion{Si}{ii}, \ion{Ca}{ii} 
and \ion{Fe}{ii} to appear in the spectrum.

The feature at 6300 \AA\ mainly consists of the \ion{Si}{ii} $\lambda$6355,
and it becomes stronger from pre-maximum epochs (Figs. \ref{fig:spec1} and 
\ref{fig:spec2}). After the maximum, Fe lines begin to contribute to this 
feature (Figs.\ref{fig:spec2} and \ref{fig:spec3}). No C lines are visible, 
giving an upper limit of C mass fraction X(C) $\lsim 0.001$. The mass fractions of 
Si, Ca, Fe and Ni can be estimated by the visible lines (Table 
\ref{tab:spparam}). The mass fraction of Si, stable Fe and \Nifs\ required for 
the fitting of this spectrum are larger than those for pre-maximum spectra.
Note again that we assume spatially homogeneous abundance distribution
for each epoch.

After maximum brightness, the evolution of the spectral shape is not
significant (Fig. \ref{fig:spec2}). Almost all the absorption features are 
identified and they are mostly \ion{O}{i}, \ion{Ca}{ii}, \ion{Fe}{ii} and 
\ion{Co}{ii} lines. The spectra are reproduced by almost the same element 
fractions (Table \ref{tab:spparam}). Estimating the mass fraction of stable 
Fe separately with \Nifs\ becomes difficult because more than $25 \%$ of \Nifs, 
which is one of the dominant elements, has decayed into \Fefs\ at $\gsim 40$ 
days after the explosion.

An interesting feature is a weak absorption at 6400 \AA\ in the spectra at day
+14 (and possibly in day +24). Branch et al. (2004) suggested 
that this is a \ion{Co}{ii} line.
However, if we reproduce this feature by Co, the line
blocking at $\lsim$ 4000 \AA\ is too strong, making the flux there too low. 
This line could be reproduced by the \ion{C}{ii} $\lambda$6578 line. This
identification is, however, quite marginal. If we assume this feature is the
\ion{C}{ii}, it requires X(C) $\sim$ 0.01. This is larger than the upper
limit derived in the preceding spectra, which traces the abundances in the 
outer layer.

The \ion{O}{i} $\lambda$7774 line is always visible in the spectra
from day +14 to + 45.
The \ion{O}{i} line is very weak in the earlier spectra,
as a result of a severe suppression of \ion{O}{i} fraction
due to the high temperature.
In fact, the modeling suggests a high O mass fraction
in the outermost layers (X(O) $\gsim$ 0.75).
These facts clearly indicate that unburned O exists in every place
of the ejecta at $v > 1000$ \kms.
The mass of O derived from spectral modeling is discussed
in \S \ref{sec:splate}.

\subsubsection{Late phase}
\label{sec:splate}

 The late phase spectra of SN 2005hk 
consist of a combination of P-Cygni profiles of permitted lines and
emission features of forbidden lines (Fig.\ref{fig:spec4}).
We note that no strong [\ion{O}{I}] line is seen in the observed spectra.
The presence of P-Cygni features suggests that
the ejecta are not completely transparent even at such late epochs.

We study the late phase spectra using the model E03, 
which has similar properties to those of 3D deflagration model.
Considering the observed spectra,
there seems to be two major problems in this model.

(1) The ejecta of the E03 model become optically thin at the late phases.
The photospheric velocity becomes zero at 250 days after the explosion
if a constant line opacity ($0.1$ ${\rm cm^2 g^{-1}}$) is assumed.

(2) In the E03 model, the mass of oxygen is as much as 0.44 $\Msun$.
Such a model would give conspicuous emission 
lines of [\ion{O}{i}] (\eg Kozma et al.\ 2005).

\begin{figure}
\plotone{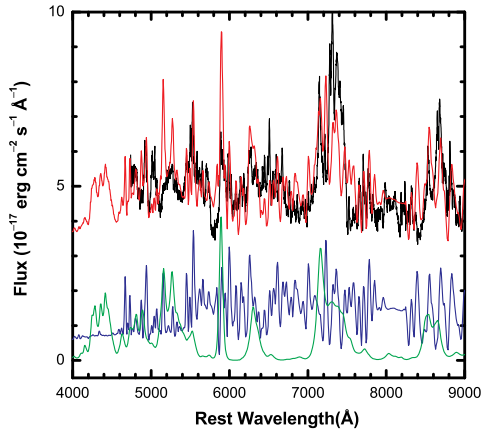}
\caption{Spectrum of SN 2005hk at +228 days with respect
to date of  $B$ maximum (black line, $3 \times 10^{-17}$
erg cm$^{-2}$ s$^{-1}$ \AA$^{-1}$ added to the flux for
presentation), as compared to
the synthetic "combined" spectrum (red line, the same amount added to
the flux). The synthetic spectrum is a combination of
the photospheric (blue line) and the nebular (green line) components.
\label{fig:nebmodel}}
\end{figure}

 First, we create synthetic spectra with the Monte Carlo code
used in the modeling of early phase spectra.
Here we assume the presence of the region emitting continuum light
at the velocity of $v_{\rm ph}  = 250$ \kms.
This could be realized if the central part of the ejecta
is denser than that of the E03 model. 
A synthetic spectrum computed with $\log$ ($L$) (\ergs) $= 40.86$
and the element mass fraction same as at day +45
(blue line in Fig. \ref{fig:nebmodel})
is in reasonable agreement with the observed spectrum at day +228
except for emission features at 7300 \AA\ ([\ion{Ca}{ii}]) and
8600 \AA\ (\ion{Ca}{ii}).
The temperature at the photosphere is as low as 4400 K.
Most of the absorption lines are \ion{Fe}{ii} but
some lines of neutral atoms such as \ion{Na}{i} and \ion{Fe}{i}
are also found.

An absorption of the \ion{O}{i} $\lambda$7774 line is seen 
as discussed in \S 6.2 (also reported in SN 2002cx, Jha et al.\ 2006).
In the synthetic spectrum, we could get the \ion{O}{i} line 
with the mass fraction X(O) $\sim 0.5$.
Therefore, we conclude that the unburned O is present
in whole ejecta, down to very low velocities, 
suggesting a nearly completely mixed abundance distribution.
 This property is consistent with 3D deflagration models.

In modeling the early and late phase spectra, the mass fractions of elements are
optimized assuming homogeneous abundance distribution above the
photosphere at each epoch. 
If we assume these mass fractions are characteristic values at each
velocity range of the ejecta, the mass of each element can be computed by
integrating the density profile of the E03 model. In Table
\ref{tab:param}, the
masses of C, O, Si and Fe group elements derived from spectral modeling are
shown. Compared with the masses of the E03 model, the smaller amount of C and
Si, and larger amount of O are preferred by the spectra. The nuclear energy
release corresponding to the element abundances obtained from the spectral
fitting is $\sim 0.8 \times 10^{51}$ erg. 
 Since the binding energy of the WD is $\sim 0.5 \times 10^{51}$erg, 
this leads to a kinetic energy of
$\sim 0.3 \times 10^{51}$ erg, which is consistent with  the E03
model.

 In the computation above, the cooling via lines is not considered,
\ie no emission lines would appear.
In order to investigate the emission lines,
we compute the late phase spectra using a nebular synthesis code
and consider the combination with the synthetic photospheric spectrum
(as in Mazzali et al. 2004).

The nebular code computes the deposition from $\gamma$-rays with a gray
opacity (0.025 cm$^{2}$ g$^{-1}$) by Monte Carlo method. 
 Positrons are assumed to be trapped in situ.
To take into account the different channels in which
the deposited energy is consumed, we assume
$L_{\rm neb} = f_{\rm neb} L_{{\gamma}, {e^+}}$
where $L_{{\gamma}, {e^+}}$ is the luminosity deposited by
the $\gamma$-rays and positrons from $0.18 \Msun$ of $^{56}$Co
(originally $^{56}$Ni).
The fraction, $f_{\rm neb}$, of $L_{{\gamma}, {e^+}}$ is
assumed to be balanced by cooling via nebular emission lines,
while the other fraction is by photospheric, allowed transitions
as modeled by the early phase spectrum synthesis code.
In other word, the luminosity (1-$f_{\rm neb}$) $L_{{\gamma}, {e^{+}}}$
is deposited below the photosphere, while the remaining luminosity
($f_{\rm neb}) L_{{\gamma}, {e^{+}}}$ is deposited into the
optically thin layer above the photosphere.
With this deposition luminosity (with varying $f_{\rm neb}$),
non-LTE rate equations
are solved assuming the ionization-recombination equilibrium
(Mazzali et al.\ 2001).
Note that the code assumes that the ejecta are optically thin.
For the calculation, we assume homogeneous distribution of
elements and $^{56}$Co. The masses of these elements are
set to be consistent with the E03 model, i.e.,
$M$($^{56}$Co) $= 0.18 \Msun$ and $M$(O) $= 0.4 \Msun$.

Figure \ref{fig:nebmodel} shows the combination of the photospheric
(as described in the first part of this section, with the luminosity
corresponding to $M$($^{56}$Co) $= 0.27\Msun$
\footnote{This luminosity is higher than that of the E03 model
because the late spectrum taken with Subaru is calibrated against
the photometry that is brighter than the model LC (Fig. \ref{fig:modelLC})
at 250 days after the explosion.})
and nebular components with $f_{\rm neb} = 0.5$.
We find that the [\ion{O}{i}]$\lambda\lambda$6300, 6363 doublet is not
strong in the model spectra. Indeed,
the resulting [\ion{O}{i}] is so weak
that it does not contradict with the non-detection of the [\ion{O}{i}],
if $f_{\rm neb} \lsim 0.5$, as is consistent with
the observed, allowed line-dominated spectrum.

This is an outcome of (a) the high density of the ejecta and
of (b) the assumption that the elements, especially O and Fe, are
homogeneously, thus microscopically, mixed together with one another.
With these two conditions, the deposited luminosity is
effectively consumed by copious [\ion{Fe}{ii}] lines
rather than [\ion{O}{i}] \footnote{Indeed, [\ion{Fe}{ii}] around 5200\AA\
is too strong in the model spectrum,
thus smaller value of $f_{\rm neb}$ is favored.}.
The high density is a direct consequence of the
low energy SN Ia model.
However, the complete mixing is a rather extreme assumption,
so that we need more detailed modeling of the late phase spectra of SN 2005hk.

With the nebular component,
we obtain better  fits compared to purely  photospheric component,
in the following aspects:
(i) Strong emission features below $\sim 5500$\AA,
(ii) [\ion{Fe}{ii}]$\lambda$7155 and [\ion{Fe}{ii}]$\lambda$7172,
(iii) [\ion{Ca}{ii}]$\lambda\lambda$7291, 7324, and
(iv) \ion{Ca}{ii} IR and [\ion{C}{i}]$\lambda$8727.

The synthetic nebular spectrum for
the E03 model results in broader lines than the observed nebular emissions.
Typical example is [\ion{Ca}{ii}]$\lambda\lambda$7291,7324.
The two components are separated in the observations, but
blended in the model. This indicates that the 
nebular line emitting region is indeed more centrally 
concentrated than the model.
This may again suggest the central part being
denser than the E03 model.

In summary, the major problems (1) and (2) above 
could be solved if the ejecta are denser than 
in the  original model and the elements are microscopically mixed.
Then, the ejecta at late phases consist of three parts: 
the optically thick innermost part 
($v \lsim 250$ \kms), 
the optically thin region emitting forbidden lines 
($v \lsim 700$ \kms, \ie the width of the [\ion{Ca}{ii}] lines), 
and the  outer layer which is thin even to $\gamma$-rays.

\section{Discussion}

We have studied the bolometric LC and early and late phase spectra 
by computing synthetic LC and spectra.
We construct a simple, less energetic model by scaling the kinetic energy 
of W7 down to $0.3 \times 10^{51}$ erg.
This model (E03) explains the LC of SN 2005hk from early to late phase.
The ejecta of the model contain as much as $\sim 0.8 \Msun$
of unburned C+O material.
In this sense, our less energetic model is similar to 
the 3D deflagration model.

Branch et al. (2004), Jha et al. (2006) and Phillips et al. (2007)
suggested that the properties of SN 2002cx-like objects are
similar to those of the 3D deflagration model.
We confirm this suggestion by modeling the LC until late phase
for the first time.
We also find that the early phase spectra of SN 2005hk 
are also explained by lower energetic model.
The \ion{O}{i} absorption is always seen until late phase,
suggesting that the elements in the ejecta are mixed almost homogeneously.
This is also consistent with the 3D deflagration model.

The late phase spectra of SN 2005hk do not show strong [\ion{O}{i}] line
and do show P-cygni profiles of permitted lines.
They seem inconsistent with what 3D deflagration model predicts, 
 at a first look.
To study this in detail, we modeled the late phase spectra of SN 2005hk 
by a combination of the photospheric and nebular components.
We find that the innermost region of the ejecta should be 
denser and more centrally concentrated 
than that of our model in order to make an optically thick region
and a smaller emitting region of nebular lines.
The absence of the [\ion{O}{i}] line is explained if 
the elements are microscopically mixed.

In summary, our less energetic explosion model or 3D deflagration
model seems to be compatible with the late phase spectrum of SN 2005hk 
(and of SN 2002cx-like objects)
(1) if the density of the central region is higher than that of 
our scaled model
and (2) if the elements are microscopically mixed,
although it is a question how such extreme conditions are realized.

The nature of this class of objects can be studied further
if observations become available at  phases later than presented here.
SN 2005hk is still not completely transparent at $\sim 400$ days
after the explosion.
Only when the ejecta become completely thin,
useful information on the structure of the innermost ejecta can be obtained.

If there is a sub-class of SNe Ia with a lower kinetic energy than in normal SNe Ia,
it could be harmful for cosmological use of SNe Ia.
To show the effect, we construct a model less energetic than 
model E03 by scaling the energy of W7 model to 
$0.08 \times 10^{51}$ erg ($\sim 1/4$ of the kinetic energy of the E03 model). 
If the deflagration is very weak and slow, such a situation will be realized.
Since the nuclear energy release is only slightly higher than 
the binding energy of the WD, such an explosion could be followed after
a pulsation (pulsating deflagration model, \eg Nomoto, Sugimoto \& Neo 1976).
Since the mass of \Nifs\ synthesized by such an explosion 
is thought to be smaller than that of the E03 model,
we set $\Mni = 0.09 \Msun$ ($1/2$ of E03).
We call this model E008, and the parameters are summarized in Table 
\ref{tab:param}.

In Figure \ref{fig:modelLC}, the synthetic LC with the E008 model 
is shown as a  magenta line.
The LC has a fainter peak magnitude than that of the E03 model 
because of the smaller amount of \Nifs.
However, the decline rate of the LC after maximum is slower 
than in the E03 model because of the lower kinetic energy of the ejecta.
This trend is opposite to the well-known LC-width relation in SNe Ia,
\ie brighter SNe Ia declines more slowly.

Such explosions can be distinguished by the late phase LC.
Since less energy leads to an efficient trapping of $\gamma$-rays
at late phases,
the decline  rate of the LC at  $\gsim 100$ days after the explosion
is slower.
This is clearly seen in the synthetic LC (Figure \ref{fig:modelLC}).
Although the peak magnitude of the E008 model 
is fainter than that of the E03 by $\sim 1$ mag, 
the late phase luminosities of these two models are comparable.

A recently discovered supernova, SN 2007qd may belong to this sub-class
(SN 2002cx-like; Bassett et al. 2007, Goobar et al. 2007).
This supernova was reported to have reached a fainter maximum than SN 2002cx 
and SN 2005hk and lower  line velocity,
being roughly half that of 2002cx or 2005hk.
Such properties are realized if the kinetic energy of the ejecta
is about quarter  that of SN 2002cx (similar to the E008 model),
because of the relation of $\KE \propto v^2$.
If this is true, the LC of SN 2007qd will evolve very slowly 
as shown in the magenta line in Figure \ref{fig:modelLC}.

\section{Conclusions}

Photometric and spectrosopic data on the peculiar SN 2005hk
are presented. The $B$ band light curve of SN 2005hk shows comparable
pre-maximum brightening, 
faster decline in the initial $\sim$20 days past maximum and
slower decline beyond $\sim$50 days past maximum, as compared to normal
Type Ia supernovae. 
The fainter peak bolometric luminosity indicates synthesis of small
amount of \Nifs \ in the explosion. Low expansion velocity of the ejecta
together with fainter peak luminosity is explained by an explosion with lower
kinetic energy. A reasonable fit to the bolometric light curve of SN
2005hk is
achieved with a less energetic ($0.30 \times 10^{51}$erg) model which
synthesized  $0.18 \Msun$ of \Nifs. 
The light curve evolution is similar to SN 2002cx.

The pre-maximum spectrum of SN 2005hk is similar to that of SN 1991T, with
much lower expansion velocity. 
The spectral evolution of SN 2005hk is very similar to SN 2002cx. 
The spectrum of SN 2005hk at late phases ($> 200$ days) is similar to 
SN 2002cx, except
for higher expansion velocities and higher velocity dispersions.
The presence of P-Cygni profiles in the late phase spectrum of SN 2005hk
indicates
that the ejecta have  not become optically thin till our last observation.
Modeling of the pre-maximum spectra of SN 2005hk indicates a relatively
higher temperature which makes \ion{Fe}{iii} lines strong.  
The presence of weak \ion{O}{i} line at $\lambda$7774 
at almost all epochs is modeled as a
consequence of high abundance of completely mixed unburned oxygen in the
ejecta.
The late phase spectra of SN 2005hk are modeled as a combination of the
photospheric and nebular components, with the nebular line emitting region
being more centrally concentrated than is expected in the case of a lower
energetic model.

\acknowledgements

We are thankful to the anonymous referee for valuable comments. This work has 
been carried out under the INSA (Indian National Science Academy) -
JSPS (Japan Society for Promotion of Science) exchange programme.
This research was supported in part by the National Science Foundation
under Grant No. PHY05-51164 and grant-in-Aid for  Scientific Research 
(18104003, 18540231) and the 21st  Century COE Program (QUEST) from 
the JSPS and MEXT of Japan. M.T. and N.T. 
are supported through the JSPS Research Fellowships for Young Scientists.
K.M. is supported through the JSPS Postdoctoral Fellowships 
for Research Abroad.
This work has used data collected at 
Subaru Telescope, which is operated by the National Astronomical 
Observatory of Japan. We thank J. Deng for useful discussions 
and S. Srividya for help in observations and data reduction.
 We thank all the observers of the
2-m HCT (operated by the Indian Institute of Astrophysics), who kindly
provided part of their observing time for the supernova observations. 
This work has made use of the NASA Astrophysics Data System and the NASA/IPAC
Extragalactic Database (NED) which is operated by Jet Propulsion Laboratory,
California Institute of Technology, under contract with the National
Aeronautics and Space Administration.


\begin{deluxetable}{lccccc}
\tablewidth{0pt}
\tablecaption{Magnitudes for the sequence of secondary standard stars in
the field of SN 2005hk.\tablenotemark{*}}
\tablehead{
ID & U  & B & V &  R & I
}
\startdata
1 &  $15.896\pm0.020$ &  $15.425\pm0.006$ & $14.643\pm0.004$ &
$14.174\pm0.012$ & $13.777\pm0.005$\\
2 &  $17.257\pm0.025$ &  $16.859\pm0.007$ & $16.067\pm0.005$ &
$15.579\pm0.010$ & $15.149\pm0.010$\\
3 &  $19.175\pm0.035$ &  $17.696\pm0.002$ & $16.375\pm0.009$ &
$15.564\pm0.007$ & $14.886\pm0.010$\\
4 &  $16.824\pm0.025$ &  $16.954\pm0.010$ & $16.449\pm0.016$ &
$16.097\pm0.016$ & $15.750\pm0.009$\\
5 &  $17.842\pm0.027$ &  $17.645\pm0.010$ & $16.902\pm0.009$ &
$16.440\pm0.018$ & $16.003\pm0.008$\\
6 &  $17.094\pm0.022$ &  $16.870\pm0.016$ & $16.150\pm0.010$ &
$15.706\pm0.017$ & $15.275\pm0.005$\\
7 &  $17.834\pm0.032$ &  $17.213\pm0.011$ & $16.210\pm0.012$ &
$15.646\pm0.022$ & $15.108\pm0.004$\\
8 &  $19.134\pm0.040$ &  $18.194\pm0.020$ & $17.175\pm0.014$ &
$16.571\pm0.013$ & $16.055\pm0.004$\\
9 &  $16.366\pm0.024$ &  $16.470\pm0.007$ & $16.115\pm0.007$ &
$15.871\pm0.012$ & $15.623\pm0.002$\\
10&  $19.516\pm0.042$ &  $18.162\pm0.020$ & $16.944\pm0.010$ &
$16.201\pm0.021$ & $15.553\pm0.001$\\
11&                   &  $18.059\pm0.008$ &                  &
$16.782\pm0.016$ &\\
\enddata
\tablenotetext{*}{The stars are identified in Figure. \ref{fig:field}}
\label{tab:calmag}
\end{deluxetable}


\begin{deluxetable}{lccccccc}
\tablewidth{0pt}
\tablecaption{Photometric observations of SN 2005hk}
\tablehead{
 Date & J.D. & Phase\tablenotemark{*} & U & B & V & R & I\\
     & 2453000+ & (days)    &   &   &   &   &
}
\startdata
05/11/2005& 680.06&  -5.22&  $              $ &$16.126\pm0.014$
&$16.167\pm0.018$& $ 16.023\pm 0.014$&  $16.060\pm 0.032$\\
06/11/2005& 681.04&  -4.25&  $15.482\pm0.046$ &$16.066\pm0.012$
&$16.047\pm0.011$& $ 15.934\pm 0.021$&  $15.979\pm 0.016$\\
07/11/2005& 682.03&  -3.27&  $15.181\pm0.086$ &$16.004\pm0.046$
&$16.007\pm0.027$& $ 15.862\pm 0.016$&  $15.866\pm 0.039$\\
08/11/2005& 683.06&  -2.25&  $15.515\pm0.060$ &$15.966\pm0.018$
&$15.900\pm0.018$& $ 15.784\pm 0.013$&  $15.798\pm 0.024$\\
13/11/2005& 688.04&   2.66&  $15.741\pm0.119$ &$16.022\pm0.069$
&$15.739\pm0.050$& $ 15.631\pm 0.034$&  $15.577\pm 0.051$\\
15/11/2005& 690.13&   4.73&  $15.956\pm0.066$ &$16.208\pm0.027$
&$15.710\pm0.024$& $ 15.564\pm 0.017$&  $15.523\pm 0.016$\\
16/11/2005& 691.20&   5.79&  $16.030\pm0.058$ &$16.260\pm0.021$
&$15.763\pm0.023$& $ 15.563\pm 0.016$&  $15.416\pm 0.018$\\
17/11/2005& 692.14&   6.72&  $16.222\pm0.062$ &$16.413\pm0.020$
&$15.779\pm0.020$& $ 15.568\pm 0.019$&  $15.466\pm 0.027$\\
19/11/2005& 694.13&   8.69&  $16.572\pm0.116$ &$16.671\pm0.021$
&$15.840\pm0.013$& $ 15.553\pm 0.011$&  $15.427\pm 0.025$\\
22/11/2005& 697.06&  11.58&  $17.212\pm0.037$ &$17.132\pm0.012$
&$16.019\pm0.008$& $ 15.649\pm 0.015$&  $15.466\pm 0.015$\\
01/12/2005& 706.06&  20.48&  $18.620\pm0.021$ &$18.260\pm0.021$
&$16.704\pm0.009$& $ 16.126\pm 0.015$&  $15.796\pm 0.019$\\
05/12/2005& 710.03&  24.40&  $              $ &$18.494\pm0.011$
&$16.925\pm0.011$& $ 16.364\pm 0.013$&  $15.970\pm 0.019$\\
08/12/2005& 713.02&  27.36&  $              $ &$18.600\pm0.019$
&$17.076\pm0.010$& $ 16.501\pm 0.017$&  $16.106\pm 0.009$\\
12/12/2005& 717.03&  31.32&  $              $ &$18.758\pm0.033$
&$17.237\pm0.024$& $ 16.646\pm 0.025$&  $16.251\pm 0.018$\\
17/12/2005& 722.06&  36.29&  $              $ &$18.832\pm0.016$
&$17.362\pm0.012$& $ 16.826\pm 0.013$&                   \\
23/12/2005& 728.13&  42.29&  $              $ &$18.846\pm0.096$
&$17.607\pm0.051$& $ 17.053\pm 0.041$&  $16.593\pm 0.044$\\
26/12/2005 & 730.79  &  44.92  &  $              $
&$18.859\pm0.010$ &                & $ 17.150\pm 0.017$&\\
26/12/2005& 731.14&  45.27&  $              $ &$18.944\pm0.024$
&$17.605\pm0.010$& $ 17.101\pm 0.009$&  $16.698\pm 0.023$\\
28/12/2005& 733.07&  47.17&  $              $ &$19.017\pm0.014$
&$17.618\pm0.013$& $ 17.158\pm 0.008$&  $16.749\pm 0.008$\\
12/01/2006& 748.06&  61.99&  $              $ &$19.204\pm0.031$
&$17.920\pm0.014$& $ 17.501\pm 0.011$&  $17.142\pm 0.027$\\
14/01/2006& 750.05&  63.96&  $              $ &$19.325\pm0.045$
&$18.038\pm0.035$& $ 17.548\pm 0.027$&  $17.203\pm 0.015$\\
30/06/2006 & 917.04 & 230.03&  $              $
&$21.578\pm0.033$ &$              $& $ 20.140\pm 0.022$& $          $  \\
04/07/2006& 921.40& 233.31&  $              $ &$              $
&$20.940\pm0.044$& $               $&  $               $\\
18/07/2006& 935.40& 247.14&  $              $ &$              $ &$
     $& $ 20.474\pm0.036$&  $               $\\
20/09/2006& 999.39& 310.38&  $              $ &$              $
&$22.733\pm0.168$& $ 21.633\pm0.099$&  $               $\\
27/11/2006&1066.73& 376.98&  $              $
&$24.206\pm0.124$ &                 & $ 21.776\pm0.036$&  $$\\
\enddata
\tablenotetext{\rlap{*}}{{ \,Observed phase with respect to the epoch of
maximum in $B$ band (JD 2453685.34), corrected for the (1+z) time-dilation
factor using the host galaxy redshift z = 0.0118}}
\label{tab:mag}
\end{deluxetable}


\begin{deluxetable}{lcrcl}
\tablewidth{0pt}
\tablecaption{Log of spectroscopic observations of SN 2005hk}
\tablehead{
Date & J.D. & \multicolumn{1}{c}{Phase\rlap{*}} & Range & Telescope\\
     & 2453000+ & \multicolumn{1}{c}{(days)} & \AA\ &
}
\startdata
04/11/05 & 679.10 & $ -6.17$   & 3500-7000; 5200-9100 & HCT\\
05/11/05 & 680.06 & $ -5.22$   & 3500-7000; 5200-9100 & HCT\\
06/11/05 & 681.06 & $ -4.23$   & 3500-7000; 5200-9100 & HCT\\
13/11/05 & 688.06 & $  2.69$   & 3500-7000; 5200-9100 & HCT\\
15/11/05 & 690.25 & $  4.85$   & 3500-7000; 5200-9100 & HCT\\
17/11/05 & 692.10 & $  6.68$   & 3500-7000; 5200-9100 & HCT\\
20/11/05 & 695.27 & $  9.91$   & 3500-7000; 5200-9100 & HCT\\
23/11/05 & 698.19 & $ 12.70$   & 3500-7000; 5200-9100 & HCT\\
24/11/05 & 699.18 & $ 13.68$   & 3500-7000; 5200-9100 & HCT\\
01/12/05 & 706.09 & $ 20.50$   & 3500-7000; 5200-9100 & HCT\\
05/12/05 & 710.08 & $ 24.45$   & 3500-7000; 5200-9100 & HCT\\
13/12/05 & 718.02 & $ 32.30$   & 3500-7000; 5200-9100 & HCT\\
26/12/05 & 730.81 & $ 44.94$    & 4000-9000 & Subaru\\
12/01/06 & 748.09 & $ 62.02$   & 3500-7000; 5200-9100 & HCT\\
30/06/06 & 916.5  & $ 228.46$  & 4700-9000 & Subaru\\
27/11/06 & 1067.05 & $ 377.26$  & 4700-9000 & Subaru\\
\enddata
\tablenotetext{*}{{ \,Observed phase with respect to the epoch of
maximum in $B$ band (JD 2453685.34), corrected for the (1+z) time-dilation
factor using the host galaxy redshift z = 0.0118}}
\label{tab:spec}
\end{deluxetable}


\begin{deluxetable}{lrrrr}
\tablecaption{Photometric parameters for SN 2005hk}
\tablehead{
Data  &  $B$  &  $V$ &  $R$ & $I$
}
\startdata
epoch of max\tablenotemark{*}  & $685.34\pm 0.4$ & $689.49\pm 0.8$ &
$691.78\pm0.2$ &$693.93\pm 0.3$ \\
magnitude at max   & $15.91\pm 0.03$ & $15.71\pm 0.04$ &
$15.55\pm 0.02$ & $15.43\pm 0.03$\\

$\Delta m_{15}(B)$  & $1.68\pm 0.05$ & \\
\\
colors at B max\tablenotemark{**} & & $B-V$ & $V-R$ & $R-I$\\
                 & & $-0.03\pm 0.04$ & $0.08\pm 0.03$ &
$-0.02\pm 0.06$\\
\\
decline rate days 20-45 (mag day$^{-1}$) & 0.021 & 0.027 & 0.031 & 0.031\\
decline rate days 230-380 (mag day$^{-1}$) & & 0.015 & 0.011 &  \\

\enddata
\tablenotetext{*}{JD 2453000+}
\tablenotetext{**}{colors are corrected for reddening
$E(B-V)_{total}=0.11$}
\label{tab:lcparam}
\end{deluxetable}


\begin{deluxetable}{lcccccc}
\tablewidth{0pt}
\tablecaption{Characteristic Quantities of the Models}
\tablehead{
Model &
$\KE$ \tablenotemark{a} &
$\Mni$ \tablenotemark{b} &
$M ({\rm C})$ \tablenotemark{c} &
$M ({\rm O})$ \tablenotemark{d} &
$M ({\rm Si})$ \tablenotemark{e} &
$M ({\rm Fe\ group})$ \tablenotemark{f}
}
\startdata
W7 (92A)   & 1.2  & 0.17  & 0.038  & 0.12  & 0.16 & 0.95\\
W7 (91bg)  & 1.2  & 0.088 & 0.038  & 0.12  & 0.16 & 0.95\\
E03        & 0.30 & 0.18  & 0.36  & 0.44  & 0.074 & 0.45 \\ 
E008       & 0.08 & 0.09  & 0.44  & 0.53  & 0.053 & 0.32 \\ \hline
Spectra    & --   & --    & $<$ 0.011  & 0.84  & 0.023 & 0.44 \\
\enddata
\tablenotetext{a}{Kinetic energy of the explosion ($10^{51}$ \ergs) }
\tablenotetext{b}{Total mass of \Nifs\ ($\Msun$)}
\tablenotetext{c}{Mass of C contained in the ejecta ($\Msun$)}
\tablenotetext{d}{Mass of O contained in the ejecta ($\Msun$)}
\tablenotetext{e}{Mass of Si contained in the ejecta ($\Msun$)}
\tablenotetext{f}{Total mass of Fe group elements including \Nifs\ ($\Msun$)}
\label{tab:param}
\end{deluxetable}


\begin{deluxetable}{ccccccccccc}
\tablewidth{0pt}
\tablecaption{Parameters of Spectral fitting}
\tablehead{
Epoch \tablenotemark{a} &
$\vph$ \tablenotemark{b} &
$M_{r, {\rm ph}}$ \tablenotemark{c} &
$\log (L)$ \tablenotemark{d} &
$X ({\rm C})$ &
$X ({\rm O})$ &
$X ({\rm Si})$ &
$X ({\rm S})$  &
$X ({\rm Ca})$ &
$X ({\rm Fe})$ \tablenotemark{e}&
$X ({\rm ^{56}Ni})$\tablenotemark{f}
}
\startdata
-6  &  6500 & 1.11 & 42.56 & $<$ 0.001 & 0.86 & 0.002 & 0.021 & 0.001  & 0.020 &
0.035  \\
+3  &  6000 & 1.05 & 42.66 & $<$ 0.001 & 0.77 & 0.020 & 0.002 & 0.001  & 0.020 &
0.14  \\
+14 &  3500 & 0.583 & 42.44 & 0.010 (?) & 0.55 & 0.020 & 0.002 & 0.004  & 0.010 &
0.39  \\
+24 &  1500 & 0.128 & 42.19 & 0.010 (?) & 0.53 & 0.020 & 0.022 & 0.004  &
\multicolumn{2}{c}{0.40\tablenotemark{ g}}  \\
+45 &  1000 & 4.90 $\times 10^{-2}$ & 41.88 & -- & 0.53 & 0.020 & 0.022 & 0.004  &
\multicolumn{2}{c}{0.40\tablenotemark{ g}}  \\
+228 & 250 & 1.18$\times 10^{-3}$  & 040.86 & -- & 0.53  & 0.020 & 0.022 & 0.004  &
\multicolumn{2}{c}{0.40\tablenotemark{ g}}  \\
\enddata
\tablenotetext{a}{Days from $B$ maximum }
\tablenotetext{b}{Photospheric velocity (\kms)}
\tablenotetext{c}{Mass coordinate at the photosphere ($\Msun$)}
\tablenotetext{d}{Bolometric luminosity (\ergs) in logarithm}
\tablenotetext{e}{Mass fraction of stable Fe}
\tablenotetext{f}{Mass fraction of radioactive \Nifs}
\tablenotetext{g}{The sum of stable Fe and $^{56}$Ni}
\label{tab:spparam}
\end{deluxetable}

\end{document}